\documentclass[aps,twocolumn,showpacs,preprintnumbers,superscriptaddress,amsmath,amssymb]{revtex4}

    \usepackage[dvips]{graphicx} 

\usepackage[english]{babel}
\usepackage{blindtext}
\usepackage{dcolumn}
\usepackage{bm}
\usepackage{ulem}
\usepackage[dvips]{color} 

\newcommand{\co}[2]{\ifcase #1 \or #2 \fi}

\newif\ifnote

\begin{document}


\title{Unusual linewidth dependence of coherent THz emission measured from
intrinsic Josephson junction stacks in the hot-spot regime}

\author{M. Y. Li}
\affiliation{Research Institute of Superconductor Electronics, Nanjing University, Nanjing 210093, China}
\affiliation{National Institute for Materials Science, Tsukuba 3050047, Japan}

\author{J. Yuan}
\email{phy.wust@gmail.com}
\affiliation{National Institute for
Materials Science, Tsukuba 3050047, Japan}

\author{N. Kinev}
\affiliation{Kotel'nikov Institute of Radio Engineering and
Electronics, Moscow 125009, Russia}

\author{J. Li}
\affiliation{National Institute for Materials Science, Tsukuba
3050047, Japan} \affiliation{Hokkaido University, Hokkaido 0600810,
Japan}

\author{B.~Gross}
\affiliation{
Physikalisches Institut and
Center for Collective Quantum Phenomena in LISA$^+$,
Universit\"{a}t T\"{u}bingen,
D-72076 T\"{u}bingen,
Germany
}
\author{S. Gu\'{e}non}
\affiliation{
Physikalisches Institut and
Center for Collective Quantum Phenomena in LISA$^+$,
Universit\"{a}t T\"{u}bingen,
D-72076 T\"{u}bingen,
Germany
}

\author{A. Ishii}
\affiliation{National Institute for Materials Science, Tsukuba 3050047, Japan}

\author{K. Hirata}
\affiliation{National Institute for Materials Science, Tsukuba
3050047, Japan}

\author{T. Hatano}
\affiliation{National Institute for Materials Science, Tsukuba 3050047, Japan}

\author{D.~Koelle}
\affiliation{
Physikalisches Institut
and
Center for Collective Quantum Phenomena in LISA$^+$,
Universit\"{a}t T\"{u}bingen,
D-72076 T\"{u}bingen,
Germany
}

\author{R. Kleiner}
\affiliation{
Physikalisches Institut
and
Center for Collective Quantum Phenomena in LISA$^+$,
Universit\"{a}t T\"{u}bingen,
D-72076 T\"{u}bingen,
Germany
}

\author{V. P. Koshelets}
\affiliation{Kotel'nikov Institute of Radio Engineering and
Electronics, Moscow 125009, Russia}

\author{H. B. Wang}
\email{hbwang1000@gmail.com} \affiliation{Research Institute of
Superconductor Electronics, Nanjing University, Nanjing 210093,
China} \affiliation{National Institute for Materials Science,
Tsukuba 3050047, Japan}

\author{P. H. Wu}
\affiliation{Research Institute of Superconductor Electronics,
Nanjing University, Nanjing 210093, China}

\date{\today}

\begin{abstract}
We report on measurements of the linewidth $\Delta f$ of THz radiation emitted from intrinsic Josephson junction stacks,
using a Nb/AlN/NbN integrated receiver for detection. Previous resolution limited measurements indicated that $\Delta f$ may be below 1 GHz -- much smaller than expected from a purely cavity-induced synchronization. While at low bias we found $\Delta f$ to be not smaller than $\sim$ 500\,MHz, at high bias, where a hotspot coexists with regions which are still superconducting, $\Delta f$ turned out to be as narrow
as 23\,MHz. We attribute this to the hotspot acting as a synchronizing element. $\Delta f$ \textit{decreases} with increasing bath temperature, a behavior reminiscent of motional narrowing in NMR or ESR, but hard to explain in standard electrodynamic models of Josephson junctions.
\end{abstract}



\maketitle

Terahertz generation utilizing stacks of intrinsic Josephson
junctions (IJJs) in the high-transition-temperature (high-T$_c$)
cuprate Bi$_2$Sr$_2$CaCu$_2$O$_{8+\delta}$ (BSCCO) has been a hot
topic in recent years, both in terms of experiment \cite{Ozyuzer07,
Wang09a, Minami09,Guenon10, Wang10a,
Tsujimoto10,Koseoglu11,Benseman11,Yamaki11, Kashiwagi12,
Tsujimoto12}
and theory
\cite{Bulaevskii07, Koshelev08, Koshelev08b, Lin08,Krasnov09,Klemm09,Hu09,Nonomura09,Tachiki09,Koyama09,Pedersen09,Grib09,Savelev10,Hu10,Lin10,Zhou10,Katterwe10, Krasnov10,Koshelev10,Tachiki11, TachikiT11, Yurgens11, Yurgens11b,Lin11b, Koyama11, Slipchenko11,Krasnov11,Asai12}.
%
Typical emission frequencies $f_p$ are 0.4 -- 1\,THz, with a maximum output power of tens of $\mu$W emitted into free space \cite{Kashiwagi12}. The IJJ stacks are usually patterned as mesas on top of BSCCO single crystals and contain 500 -- 2000 IJJs. A large fraction of these IJJs oscillate coherently.
There are good indications that the synchronization has to do with electric fields utilizing the mesa as a cavity: (i) $f_p$ increases inversely proportional to the junction width \cite{Ozyuzer07,Kashiwagi12} or fulfills the resonance condition for cylindrical mesas \cite{Tsujimoto10,Guenon10}, (ii) Low-temperature scanning-laser microscopy imaged standing waves at bias points close to THz emission maxima \cite{Wang09a,Wang10a,Guenon10}.
On the other hand, (iii) the linewidth $\Delta f$ of THz radiation -- so far mostly resolution limited to some GHz
-- is much lower than the estimated linewidth of the cavity resonances \cite{Wang10a,Krasnov11} and thus one cannot simply assume that the IJJs are slaved by the cavity; (iv) in some cases THz emission lines were fairly tunable or they occurred at positions not compatible with cavity resonances \cite{Kashiwagi12}.
A recent mixing experiment \cite{Kashiwagi12} has revealed a $\Delta f$ as low as $\sim$ 0.5\,GHz at low bias, rising even more questions about the mechanism of synchronization.


In the IJJ stacks, operated at a bath temperature $T_b$ well below $T_c$, there are two emission regimes. At moderate input power (``low-bias regime'') there is only little heating, and the temperature distribution in the mesa is roughly homogeneous and close to $T_b$. At high input power (``high-bias regime'') a hotspot (an area heated to above $T_c$) forms inside the mesa, leaving the ``cold'' part of the mesa for THz generation by the Josephson effect. The hotspot seems to take part in synchronization \cite{Wang10a,Guenon10,Yurgens11,Tachiki11}.


\begin{figure}[tb]
\includegraphics[width=0.8\columnwidth,clip]{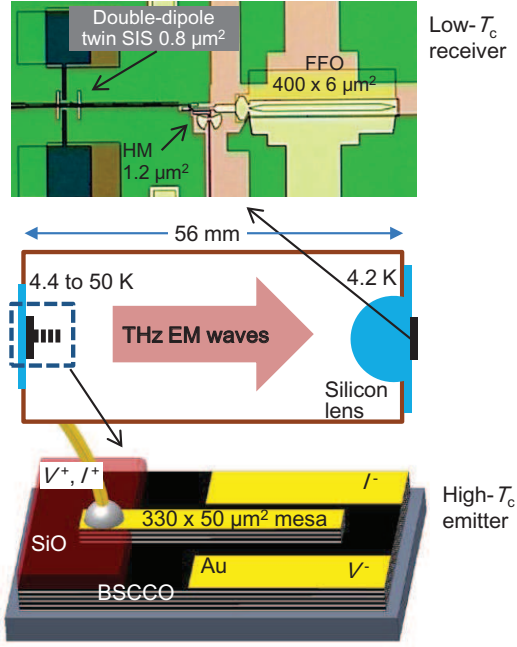}
\caption{(color online).
Setup for detection of THz emission from intrinsic Josephson junctions (sketched at bottom).
Top: central part of the superconducting integrated receiver (SIR) chip with a double-dipole antenna, a twin SIS-mixer and a harmonic mixer (HM) for flux-flow oscillator (FFO) phase-locking. For thermal shielding an infrared filter (2\,mm thick GoreTex film) was installed between the BSCCO oscillator and the SIR.
}
\label{fig:Setup}
\end{figure}
%

To measure $\Delta f$
and thus to gain more insight into the mechanism of synchronization, we have used a Nb/AlN/NbN superconducting integrated receiver (SIR)\cite{Koshelets00} for detecting THz emission from an IJJ mesa containing $\sim$ 600 junctions, cf.
Fig. \ref{fig:Setup}.

The $330 \times 50\,\mu$m$^2$ wide and 0.9\,$\mu$m thick mesa was patterned on a BSCCO single crystal ($T_c \approx$  82\,K), as described elsewhere \cite{Wang09a,Wang10a,Guenon10}.
Mesa and base crystal were contacted by Au wires fixed with silver paste.
Contact resistances ($\sim$ 1\,$\Omega$) have been subtracted in the data discussed below.
``Conventional'' THz emission measurements were performed in Tsukuba, using a Fourier spectrometer and a Si bolometer, as described in \cite{Wang10a}.
Measurements with the SIR were performed in Moscow. The SIR comprises on a single chip a planar antenna combined with a superconductor-insulator-superconductor (SIS) mixer, a superconducting flux-flow oscillator (FFO) acting as local oscillator (LO) and a SIS harmonic mixer (HM) for the FFO phase locking.
The  frequency of the LO is continuously tunable from 350 to 750\,GHz, while the SIS mixer is effectively matched to the planar antenna between 450 and 700\,GHz, limiting the SIR operation range. A frequency resolution of the SIR well below 100 kHz has been confirmed \cite{Koshelets10}.

\begin{figure}[tb]
\includegraphics[width=\columnwidth,clip]{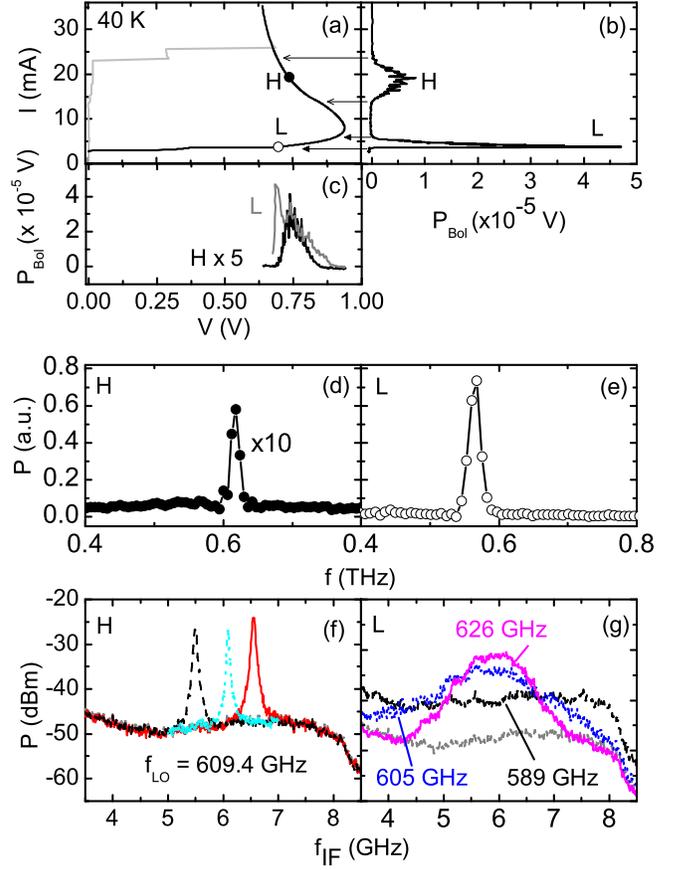}
\caption{(color online).
Results of the THz emission experiment at 40\,K: (a) IVC of the IJJ stack.
 THz emission power vs. current (b) and voltage (c).
(d), (e) Fourier spectra of emitted radiation taken at points H and
L in (a).
Spectra of the mixing output of the SIR for (f) three bias points in the high-bias regime  and
(g) for three bias points at low-bias. In (g) $f_{\rm{LO}}$ was adjusted for each bias to bring the IF signal to the center of the spectrum analyzer. $f_{\rm{LO}}$ values are indicated in the graph. Gray line indicates background noise.
}
\label{fig:Emission}
\end{figure}
%
Fig. \ref{fig:Emission} shows a typical set of data at $T_b$ = 40\,K. The current-voltage curve (IVC) of the mesa is shown in Fig. \ref{fig:Emission}(a). Upon increasing bias current $I$, near $I \sim$ 25\,mA,  one observes jumps (grey line of IVC) where the IJJs switch into their resistive state. When lowering $I$ from $\sim$ 40\,mA (black line) all junctions are resistive. Switch-back occurs below  3.65\,mA.  The outermost branch, where the emission data are taken, exhibits the typical back-bending arising from strong heating \cite{Yurgens11}. The high-bias regime embraces the region of negative differential resistance on this branch.
Figs. \ref{fig:Emission}(b),(c) show the broadband THz emission power $P_{\rm{bol}}$, simultaneously detected with the IVC by a bolometer,  vs. (b) current $I$ and (c) voltage $V$ across the mesa. In (b) one observes a broad (with respect to $I$) emission peak in the high-bias regime and a sharper one at low bias which, for this particular mesa, was more intense than the high-bias peak.
$P_{\rm{bol}}$ vs. $V$, Fig. \ref{fig:Emission}(c), shows a smooth peak for the high-bias regime, with a full-width-at-half-maximum in voltage of about 75\,mV (black line).
By contrast the low-bias signal (grey line) seems to exhibit structure, indicating different groups of oscillating IJJs.
Emission spectra, measured using a Fourier spectrometer, are shown
in Figs. \ref{fig:Emission}(d), (e).  Spectrum (d) has been taken at
point H in (a) where $V$ = 736\,mV, $I$ = 19.43\,mA. The emission
peak is at $f_p$ = 618\,GHz, corresponding to $N=V/(f_p\Phi_0)
\approx$ 576 oscillating IJJs. $\Phi_0$ is the flux quantum. The
(resolution limited) linewidth of the emission peak is $\sim
15$\,GHz. Fig. \ref{fig:Emission} (e) shows a  spectrum for bias
point L where $V$ = 697\,mV, $I$ = 3.76\,mA. The emission peak,
having a width which is also near the resolution limit, is at $f_p$
= 594\,GHz, corresponding to $N \sim$ 567.

The SIR measurements,
 showing the emitted radiation at the IF frequency, were performed near points H (Fig. \ref{fig:Emission}(f)) and L (Fig. \ref{fig:Emission}(g)). In Fig. \ref{fig:Emission}(f) the LO frequency of the SIR is $f_{\rm{LO}}$ = 609.4\,GHz. We show three measurements made for slightly ($\pm$ 0.6\,mV) different values of $V$. There is a single emission peak which shifts with increasing voltage. 
$\Delta f$ of the two outermost peaks is $\sim$ 60 MHz, while the center one is even sharper, $\Delta f \sim$ 40 MHz.
By contrast, the linewidths seen in the low bias regime, cf. Fig. \ref{fig:Emission} (g), are much larger and often exceeded the 6 GHz bandwidth of the IF amplifier. The lowest values  are $\sim$ 0.5\,GHz, in agreement with previous measurements \cite{Kashiwagi12}.

\begin{figure}[tb]
\includegraphics[width=0.9\columnwidth,clip]{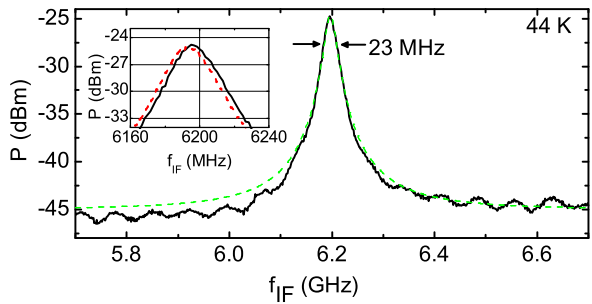}
\caption{(color online).
Spectrum (most narrow emission line obtained) of the mixing output of the SIR for $I$ = 16.3\,mA, $V$ = 0.721\,V in the high-bias regime at $T_b$ = 44\,K.  The LO frequency is $f_{\rm{LO}}$ = 605.75\,GHz; $f_{\rm{p,IF}}$ = 6.2 GHz. Dashed line is a Lorenzian fit with $\Delta f$ = 23\,MHz. Inset shows drift of emission peak within 3 minutes.
}
\label{fig:best}
\end{figure}

Fig. \ref{fig:best} shows the most narrow line ($\Delta f$ = 23\,MHz) we have obtained in the high-bias regime. Data were taken at $T_b$ = 44\,K, $I$ = 16.3\,mA and $V$ = 0.721\,V. From
$f_{\rm{LO}}$ = 605.75\,GHz and $f_{\rm{p,IF}}$ = 6.2 GHz we find $f_{\rm{p}}$ = 611.95\,GHz and $N \approx$ 570.
This value for $\Delta f$ is already quite close to the limit value that can be phase-locked by a regular room temperature semiconductor phase-locked-loop (PLL) system with regulation bandwidth of about 20\,MHz, presently limited  by the delay in the cable between the oscillator inside the cryostat and the PLL system (this limitation can be overridden by use of a cryogenic PLL system with regulation bandwidth $>$ 40\,MHz \cite{Khudchenko09}).
We also investigated the stability of the THz emission line. The inset of Fig. \ref{fig:best} shows two measurements, taken at a 3 min. interval. The 3\,MHz drift observed was smooth and unidirectional
and even lower than the typical 3 min. drift ($\sim$ 10\,MHz) of a free-running Nb based FFO.
%
We also note that the emission lines observed at high bias have Lorentzian shape, cf. dashed line in Fig. \ref{fig:best}. Recent 3D simulations \cite{Asai12} based on the sine-Gordon equation and taking a hotspot into account yielded an asymmetric Fano shape in contrast to our experimental data.

If $\Delta f$ of the high-bias THz emission signal were set by a cavity resonance, its quality factor  $Q = f/\Delta f$ would have to amount some $10^4$, which is more than unrealistic for an IJJ stack \cite{Katterwe10,Krasnov11,Katterwe11}. At least for small mesas, cavity resonances become overdamped for T $>$ 60\,K \cite{Katterwe11}.
For $T \sim$ 40\,K we expect $Q << 50$.
%
The low value of $\Delta f$ implies, that, when changing the bias current in Fig. \ref{fig:Emission}(c), a sharp emission line ``moves'' through the $\Delta V$ = 75\,mV wide emission peak. The ratio $V/\Delta V \sim 10$ of this peak would indeed be compatible with the notion of a cavity resonance.

\begin{figure}[tb]
\includegraphics[width=0.9\columnwidth,clip]{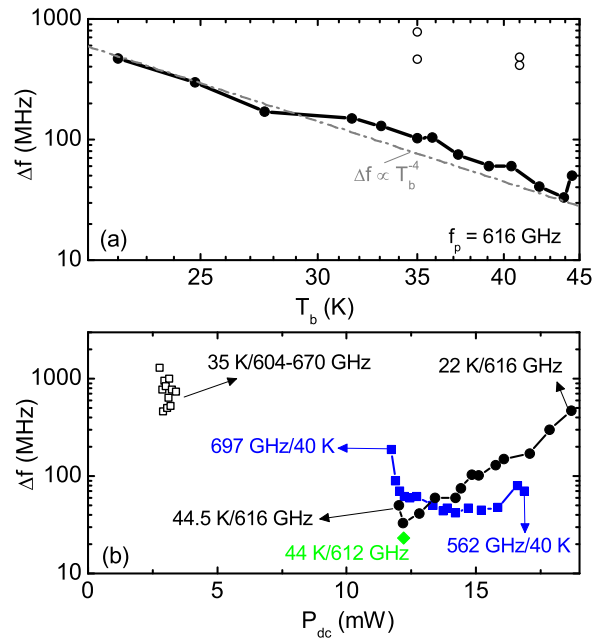}
\caption{(color online).
Linewidth vs. (a) bath temperature and (b) dc input power. In (a) solid (open) circles are for high (low) bias. In (b) open squares are for low bias, other symbols for high bias. Solid (black) circles: same data as in (a). Solid (blue) squares: linewidth at $T_b$ = 40$\pm$1\,K and various values of $f_p$. Linewidth of Fig. \ref{fig:best} is shown by solid (green) diamond.
}
\label{fig:differential}
\end{figure}


In Fig. \ref{fig:differential} (a) we show by black circles $\Delta f$ vs. $T_b$. Strikingly, $\Delta f$ \textit{decreases} with increasing $T_b$ by about a factor 15 and, roughly, $\Delta f \propto T_b^{-4}$, cf. grey line in Fig. \ref{fig:differential}(a).
For comparison we have also added for two temperatures low-bias data for $f_p$ near 616\,GHz. In this regime it was difficult to obtain systematic data points, because at most temperatures only a broadband ($>$ 6\,GHz) signal appeared in the SIR output. This further indicates that at low bias phase-lock is incomplete.
Noting that $T_b$ may not be the best variable for our system, in Fig. \ref{fig:differential} (b) we also show $\Delta f$ vs. the dc input power $P_{\rm{dc}}=IV$, where for $V$ we have used the measured value, not corrected for contact resistance. Low-bias data are shown by open squares. The black circles show the same high-bias data as Fig. \ref{fig:differential}(a); $\Delta f$ increases with increasing $P_{\rm{dc}}$.
Note that the smallest $P_{\rm{dc}}$ is at high $T_b$ and the highest $P_{\rm{dc}}$ is at the lowest $T_b$.
This does, however, not imply that the sample is hotter at low $T_b$. Simulations of the temperature distribution, performed analogous to the calculations in \cite{Yurgens11}, yielded temperatures of the cold part of the mesa which are some 10--15\,K above $T_b$, but still increase with increasing $T_b$. For the hot part we find peak temperatures $T_h$ in the range 80--110\,K, with the highest value indeed at the lowest value of $T_b$.
For comparison we show by solid squares in Fig. \ref{fig:differential}(b) $\Delta f$ out of a measurement series, where we have fixed $T_b$ to 40$\pm$1\,K and varied $f_p$. Within this curve $f_p$ varies monotonically from 697\,GHz to 562\,GHz. In contrast to the $f_p = const.$ curve, $\Delta f$ is about constant at intermediate values of $P_{\rm{dc}}$ and increases outside of this regime. Thus there seems to be no systematic dependence of $\Delta f$ on $P_{\rm{dc}}$.
%
Further note, that a slight increase of $\Delta f$ at low $P_{\rm{dc}}$ is visible both on the $T_b = const.$  and the $f_P = const.$ curve. Here, the bias is close to the hotspot nucleation point and it is possible that the hot part of the stack already is close to or even below $T_c$, \textit{i.e.} we approach the low-bias regime.

We next compare the high-bias data to theoretical predictions. We decompose $\Delta f$ into a sum of two contributions $\Delta f_T$ and $\Delta f_x$, as it has been done in an analysis of the radiation linewidth of the FFO \cite{Golubov96}. $\Delta f_T$ arises from fluctuations of the quasiparticle current (Nyquist noise) and $\Delta f_x$ represents additional fluctuations of -- at this point -- unspecified origin. For a current biased lumped junction $\Delta f_T = (4\pi/\Phi_0^2)(R_d^2/R)k_BT$, where $R$ is the junction resistance, $R_d$ is the differential resistance of the IVC at the bias point and $T$ is the junction temperature \cite{Dahm69,Jain84}. For a phase-locked 1D array of  $N$  junctions the linewidth is expected to decrease $\propto$ $N^{-1}$ \cite{Jain84}. In our case, in the high-bias regime the hot part of the IJJ stack is an excellent candidate for a shunting network causing phase-locking \cite{Guenon10,Wang10a,Yurgens11,Tachiki11}.
Our IJJ stacks are not lumped and also $T$ not only differs from $T_b$ but also varies strongly within the stack. Nonetheless an analysis of $\Delta f_T$ may give some indication whether or not $\Delta f$ vs. $T_b$ can be explained via Nyquist noise. We first look at the cold part of the mesa only. 
The out-of-plane resistance $R_c$ of BSCCO has a negative temperature coefficient (roughly $R_c \propto T^{-1}$) and thus  $T/R \propto T^2$.
We cannot infer $R_d$ from the IVC, since $T$ varies strongly along this curve. However, from small mesas it is known that $R_d$ is smaller but not very different from $R$.
Thus, an unrealistic $R_d \propto T^{-3}$ would be required to explain the data.
We also considered a hot resistor (resistance $R_s$, temperature $T_s$), representing the hotspot, in parallel to a cold one (resistance $R_J$, temperature $T_J$), as it has been analyzed in \cite{Larkin68} for a single junction,
with the result that $R$ should be replaced by $R_{\rm{eff}} =( R_s^{-1}+R_J^{-1})^{-1}$ and $T$  by $T_{\rm{eff}} = T_s/R_s + T_J/R_J$. $R_{\rm{eff}}$ follows from the measured IVC. With $T_s \approx T_h >> T_J$ and $R_s << R_J$
we get $T_{\rm{eff}} \approx T_h$ and, with $R_d \approx R_{\rm{eff}}$, one obtains an almost temperature independent $\Delta f_T \approx$ 250\,MHz. Dividing this by $N$ yields a linewidth which not only has the wrong temperature dependence but also is much lower than the measured $\Delta f$.

The second contribution to $\Delta f$, $\Delta f_x$, may arise from moving fluxons. This is the case for the FFO where, however, $\Delta f_x \propto T^{1/2}$ was found \cite{Golubov96}. Recently, THz generation in IJJ stacks by bound fluxon-antifluxon pairs (breathers) oscillating near a cavity resonance was proposed. Whether or not this mechanism leads to the observed $T$ depencence of $\Delta f$ is unclear, although we would expect an increase with $T$, as for the case of the FFO. Also, more conventional simulations based on 1D-coupled-sine-Gordon equations yielded a $\Delta f$ which increased with temperature \cite{TachikiT11}.

Another hint may be, that at low $T_b$ the in-plane thermal gradient in the stack is larger than at high $T_b$. However, we cannot translate this to the observed  $\Delta f$ vs. $T_b$ dependence.

Let us thus also look at other effects in physics that cause a
decrease of $\Delta f$ with increasing temperature. Motional
narrowing in NMR or ESR are prominent examples
\cite{Kumar66,Hendrickson73}. It is tempting to speculate about an
analogous effect in the IJJ stacks. In a nutshell, motional
narrowing occurs when some spatially varying field, causing line
broadening, is reduced on average due to temporal fluctuations. In
our case this could be due to a time-dependent hotspot. Generally,
it is possible that the hotspot oscillates \cite{Gurevich87}. We
have searched for such oscillations for frequencies in the kHz and
MHz range but could not find any. Still, the temperatures of the hot
and cold parts of the stack may fluctuate, causing fluctuations of
the hotspot edge position and as a result also variations in
junction parameters such as critical current and resistance. Suppose
that the static spread in these parameters is too large to achieve
complete phase-lock. Temporally fluctuating parameters could lead to
a situation, where for some time these parameters are in the
synchronization window, causing improved phase-locking.

To conclude:
At low bias the best values for $\Delta f$ are on the order of 0.5\,GHz, consistent with previous measurements \cite{Kashiwagi12}. However, the irregular lineshape and the strongly varying linewidth indicate that not all junctions are phase-locked and a superposition of several lines is observed.
In the high-bias regime we observe much lower values of $\Delta f$ down to $\sim$23\,MHz.
The improved linewidth compared to the low-bias regime suggests that the hotspot acts as a shunting element providing phase-lock.
$\Delta f$, measured at fixed emission frequency, \textit{decreases} with increasing temperature, reminiscent on motional narrowing observed in NMR or ESR.  This temperature dependence is hard to explain within
standard models for the linewidth of oscillating Josephson junctions.
In any case, the narrow emission line observed in the high-bias regime  makes intrinsic Josephson junction stacks interesting for, \textit{e.g.}, THz spectroscopy.

\acknowledgments
We gratefully acknowledge financial support by the JST/DFG strategic Japanese-German International Cooperative Program, the Grants-in-Aid for scientific research from JSPS
and RFBR projects 12-02-00882, 11-02-12195-ofi-m, 11-02-12213-ofi-m, Grants 2456.2012.2 and 02.740.11.0795.


%
%

\end{document}